\documentclass[a4paper,10pt]{article}
\usepackage[position=t,singlelinecheck=off]{subfig}
\usepackage{amssymb,amsmath,graphics,graphicx,float,braket}
\usepackage[colorlinks=true, citecolor=blue, urlcolor=blue ]{hyperref}
\usepackage{setspace}
\usepackage{cite}
\usepackage{mathrsfs}
\usepackage{bm,nicefrac}
\usepackage[title]{appendix}
\def\beq{\begin{equation}}
\def\eeq{\end{equation}}
\def\bea{\begin{eqnarray}}
\def\eea{\end{eqnarray}}
\def\nn{\nonumber}

\makeatletter
\def\@cite#1#2{${\mbox{#1\if@tempswa , #2\fi}}$}
\makeatother

\renewcommand{\thesection}{\Roman{section}}

\newcommand*{\footnotemarkcolor}{black}
\makeatletter
\renewcommand*{\@makefnmark}{\hbox{\@textsuperscript{%
			\color{\footnotemarkcolor}\normalfont\@thefnmark}}}
\makeatother
\usepackage{authblk}
\usepackage[symbol]{footmisc}

\def\correspondingauthor{\footnote[2]{Corresponding author: prosenjit.maity@rkmrc.in, orcid id:  0000-0001-5417-8654.}}

\begin{document}
\thispagestyle{empty}
\begin{center}
\begin{LARGE}
\textsf{Adiabatic elimination in the presence of multiphoton transitions in atoms inside a cavity}
\end{LARGE} \\

\bigskip\bigskip
Prosenjit Maity \correspondingauthor{}
\\
\begin{small}
\bigskip
\textit{
 Department of Physics, Ramakrishna Mission Residential College, \\ Narendrapur, Kolkata-700103, India.}
\end{small}
\end{center}

\vfill
\begin{abstract}
Various approaches have been used in the literature for eliminating nonresonant levels in atomic systems and deriving effective Hamiltonians. Important among these are elimination techniques at the level of probability  amplitudes, operator techniques to project the dynamics on to the subspace of resonant levels, Green's function techniques, the James' effective Hamiltonian approach, etc. None of the previous approaches is suitable for deriving effective Hamiltonians  in intracavity situations. However, the James' approach does work in the case of only two-photon transitions in a cavity. A generalization of the James' approach works in the case of three-photon transitions in a cavity, but only under Raman-like resonant conditions. Another important approach for adiabatic elimination is based on an adaptation of the Markov approximation well-known in the theory of system-bath interactions. However, this approach has not been shown to work in intracavity situations. In this paper, we present a method of adiabatic elimination for atoms inside cavities in the presence of multiphoton transitions. We work in the Heisenberg picture, and our approach has the advantage that it allows one to derive effective Hamiltonians even when Raman-like resonance conditions do not hold.

\end{abstract}

\newpage
\setcounter{page}{1}

\section{Introduction}
Various approaches for the adiabatic elimination of nonresonant levels in atomic systems to derive effective Hamiltonians have been studied in the literature.  Several of these approaches are inspired by existing approaches in the theory of system-bath interactions [\cite{gerry1,cardimona,gerry2,gerry3,wang1992,you2003,brion2007,paulisch2014,ma2015}]. Such effective Hamiltonians have played an important role in the study of atom-field interactions, both in free space as well as intracavity situations [\cite{gerry1}]. A well-known example is that a three-level lambda system, where an effective Hamiltonian for the Raman transition is derived by adiabatically eliminating the intermediate nonresonant level.

\par

We shall briefly review here some of these approaches. The simplest ones are based on adiabatic elimination of nonresonant levels in terms of the relevant probability amplitudes. More sophisticated approaches employ a combination of projection operator and Green's function techniques. While these approaches work very well in the case of few-level systems, they become cumbersome when a large number of levels are involved. In particular, in intracavity situations, where the cavity modes necessarily span an infinite dimensional Hilbert space the above-mentioned methods are difficult to adopt. An important alternative to the above-mentioned approaches was given James and Jerke [\cite{james2}]. They have given a remarkably simple derivation of an effective Hamiltonian for a multilevel system in the presence of two-photon processes, using second-order perturbation theory. Although James and Jerke did not design their approach to work in intracavity situations, it turns out that their approach as has been shown by P. Z. Zhao et al [\cite{zhao2018}], does allow one to derive an effective Hamiltonian in intracavity situations as well. 

\par

W. Shao et al [\cite{gjames}] have given a generalization of the James' approach to the case of three-photon transitions by using third-order perturbation theory. Again, W. Shao et al did not indicate whether their approach would work  in intracavity situations. In the spirit of the analysis of Zhao et al, we have shown in this paper, that their approach indeed works in intracavity situations as well. However, it is possible to derive an effective Hamiltonian only in the case of a Raman-like resonance condition. The generalized James' approach fails in situations where such a resonant condition is not satisfied. Recently, V. Paulisch et al [\cite{paulisch2014}] have presented a method of adiabatically eliminating irrelevant nonresonantly coupled levels by employing a Markov approximation on the integro-differential equation for the probability amplitudes of the relevant levels. They have also discussed a way of going beyond adiabatic elimination in terms of a hierarchy of approximations. 

\par

The aim of the present paper is to present an adaptation of the Markovian approach to derive effective Hamiltonians for atoms interacting with fields in the presence of multiphoton transitions and in cavities. An important difference between the approach of Paulisch et al and ours is that we work in the Heisenberg picture. This allows us to treat the dynamics of the atoms as well as cavity modes in a similar manner. The effective Hamiltonians that can be derived using our approach do not require any conditions such as Raman-like resonance to be satisfied.

\par

This paper is organized as follows. In Sec. \ref{review}, we briefly review some of the approaches for deriving  effective Hamiltonians which are relevant in our context. At first, we discuss the simplest one among them based on the analysis of Brion et al [\cite{brion2007}] in the context of Raman transition in a three-level lambda system interacting with  classical laser fields. Thereafter, we  review more sophisticated approach, a combination of projection operator and Green's function techniques. Subsequently, we revisit the James'  approach and its generalization [\cite{james2,gjames}] and discuss their merits and demerits in the case of multiphoton transitions in intracavity situations.  Furthermore, we  review the integro-differential equation approach of Paulisch et al [\cite{paulisch2014}] and discuss its adaptability in the context of intracavity situations. In Sec. \ref{Markov_approx}, we  introduce our approach i.e. adiabatic elimination in the Heisenberg picture using Markov approximation for multiphoton transitions inside a cavity. We demonstrate it  with a few examples and compare the effective Hamiltonians which are derived using James' and generalized James' techniques in the subsequent sections.  Section \ref{con} concludes the paper with a brief summary and some discussion. 

\section{A brief review of adiabatic elimination methods }
\label{review}

\subsection{Simplest technique of adiabatic elimination using Brion et al approach}
\label{simplest}
\begin{figure}
	\begin{center}
		\captionsetup[subfigure]{labelformat=empty}
		\includegraphics[width=5.2cm,height=4.5cm]{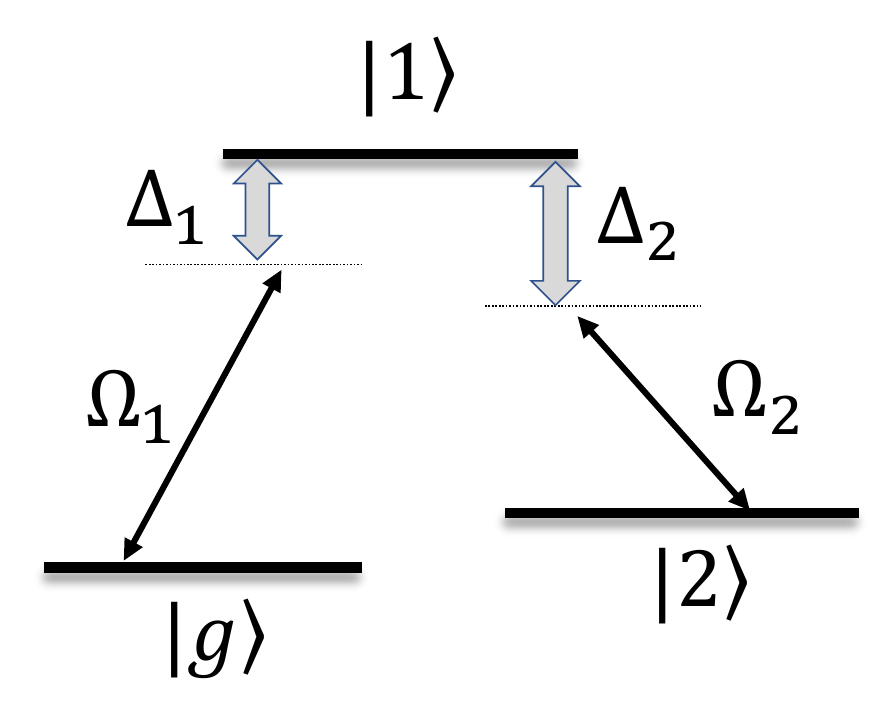} 
		\caption{Energy-level diagram of the three-level atom in lambda configuration. The detuning reads: $\Delta_{1} = \omega_{1g} - \omega_{g}^{(1)}; \; \Delta_{2} = \omega_{12} - \omega_{1}^{(l)}$. The transition between $\ket{g}$ and $\ket{1}$ is governed by Rabi frequency $\Omega_{1}(t)$ whereas the transition between $\ket{1}$ and $\ket{2}$ is controlled by the Rabi frequency $\Omega_{2}(t)$. } 
		\label{sim_Rama}
	\end{center}
\end{figure}

To begin with, we adopt a simple example to describe the adiabatic elimination technique. We look at an atomic lambda system [\cite{brion2007}]  with two lower states $\ket{g}$, $\ket{2}$ and an excited state $\ket{1}$ connected by two off-resonance lasers. The transition $\ket{g} \leftrightarrow \ket{1}$ is governed by the classical laser pulse having Rabi frequency $\Omega_{1}(t)$ and the detuning $\Delta_{1}$, whereas the transition $\ket{1} \leftrightarrow \ket{2}$ is facilitated by the Rabi frequency $\Omega_{2}(t)$ and the detuning $\Delta_{2}$, as shown in the Fig. (\ref{sim_Rama}). The detunings are read as: $\Delta_{1}  =  (\omega_{1} - \omega_{g})  - \omega_{g}^{(l)};$\;  and \;  $\Delta_{2}  =  (\omega_{1} - \omega_{2})  - \omega_{1}^{(l)}$, where $\omega_{g}^{(l)}$ and $\omega_{1}^{(l)}$ are  frequencies of the laser pulses. The Hamiltonian describing the system can be written as $(\hbar  = 1\;  \text{herein})$
\bea
\mathcal{H} &=&  \omega_{g} \sigma_{gg}  + \omega_{1} \sigma_{11} + \omega_{2} \sigma_{22}  + \Omega_{1} \exp{(- i \omega_{g}^{(l)} t)} \;\sigma_{1g}  +  \Omega_{2} \exp{(- i \omega_{1}^{(l)} t)} \;\sigma_{12} + \text{H.c}.,
\label{simplest_Raman}
\eea
where the atomic transition operators  $\sigma_{ij} \equiv \ket{i} \bra{j}$.  In the rotating frame defined by the transformation $\ket{\psi} \rightarrow \ket{\chi} = \exp{(i \zeta t)} \ket{\psi}$, where
\beq
\zeta = 
\begin{pmatrix}
\frac{\Delta}{2} &  0 & 0\\
0  &  \omega_{g} -\omega_{1} - \frac{\Delta}{2} & 0\\
0 & 0 & \omega_{g} - \bar{\Delta}	
\end{pmatrix},
\eeq
and carrying out the rotating wave approximation, the Hamiltonian (\ref{simplest_Raman}) in the unperturbed energy basis $\{\ket{g}, \ket{2}, \ket{1}\}$ reads as

\beq
\mathcal{H}^{'} = 
\begin{pmatrix}
	-\frac{\Delta}{2} &  0 & \frac{\Omega_{1}^{*}}{2}\\
	0  &  \frac{\Delta}{2} &  \frac{\Omega^{*}_{2}}{2}\\
	\frac{\Omega_{1}}{2} &  \frac{\Omega_{2}}{2} & \bar{\Delta}
\end{pmatrix},
\label{ms_raman}
\eeq
 where the abbreviations:  $\Delta  = \Delta_{1}- \Delta_{2}$  and  $\bar{\Delta}  = \frac{1}{2}(\Delta_{1}+ \Delta_{2})$. 

\par

To establish a two-photon Raman transition between the states  $ \ket{g}$ and $\ket{2}$, we adiabatically eliminate the nonresonant level $\ket{1}$ under the assumption of large detuning  condition i.e. $|\Delta|, |\bar{\Delta}| \gg \Omega_{1}, \Omega_{2}$. The most common method for removing the excited state by invoking the  Schr{\"o}dinger equation written for the state $\ket{\psi} = C_{g} \ket{g} + C_{1} \ket{1} + C_{2} \ket{2}$,
\bea
\begin{cases}
	i \dot{C}_{g} (t)  \;\; = \; \; - \frac{\Delta}{2} C_{g} + \frac{\Omega^{*}_{1}}{2} C_{1}   \\
	i \dot{C}_{1} (t) \;\;  = \; \;   \frac{\Omega_{2}}{2} C_{g} + \frac{\Omega_{1}}{2} C_{2} + \bar {\Delta} C_{1},\\
	i \dot{C}_{2} (t)  \; \; =\; \;   \frac{\Delta}{2} C_{2} + \frac{\Omega^{*}_{2}}{2} C_{1}.
\end{cases}
\label{sch_eq}
\eea
 Since  under the  large detuning condition,  the excited state $\ket{1}$ is nonresonantly coupled to the states $\ket{g}$  and $\ket{2}$,  we can claim $\dot{C}_{1} (t)  = 0$, which implies
 \beq
C_{1}  = -\frac{\Omega_{1}}{2 \bar{\Delta}} C_{g}   - \frac{\Omega_{2}}{2\bar{\Delta}} C_{2}.
\label{sol_excited}
 \eeq
  
Inserting the result in (\ref{sol_excited}) into the dynamical equations of $C_{g}$ and $C_{2}$ in (\ref{sch_eq}), we have
\beq
i
\begin{pmatrix}
	\dot{C}_{g}\\
	\dot{C}_{2}
\end{pmatrix}
= \mathcal{H}_{\text{eff}} 
\begin{pmatrix}
	C_{g}\\
	C_{2}
\end{pmatrix}, 
\eeq
where the effective Hamiltonian describing the two-photon Raman transition reads,

\beq
\mathcal{H}_{\text{eff}}  \;\; = \;\; - 
\begin{pmatrix}
	 \frac{\Delta}{2}  + \frac{|\Omega_{1}|^{2}}{4 \bar{\Delta}}  & \frac{\Omega_{1}^{*} \Omega_{2}}{4 \bar{ \Delta}}\\
	 
	 \frac{\Omega_{1} \Omega_{2}^{*}}{4 \bar{\Delta}} &    - \frac{\Delta}{2}  + \frac{|\Omega_{2}|^{2}}{4 \bar{\Delta}}
\end{pmatrix}.
\eeq

\vskip 0.3 in

\subsection{ Green’s function and projection operator formalism }
\label{projection}
The  Green’s function and projection operator formalism was first introduced in the Refs. [\cite{proj1, proj2}] and was extensively presented in the Ref. [\cite{cohenbook}]. For a given  system of  Hamiltonian $H $ which  is consisting of two parts, the  unperturbed part $H_{0}$ and a small perturbed part  $H^{'}$, one can define the Green's function $\mathcal{G}(\xi)$, $\xi \in \mathbb{C}$ as follows: 
 \beq
 \mathcal{G}(\xi)  = \frac{1}{\xi-H}.
 \label{green}
 \eeq

Let $\mathcal{P}$ be projector onto a subspace $\mathscr{E}_{0}$  subtended by an ensemble unperturbed states  $\{\ket{\phi_{1}, \ket{\phi_{2}},....,\ket{\phi_{N}}}\}$ of $H_{0}$. Assuming  $\ket{\phi_{1}},\ket{\phi_{2}},....,\ket{\phi_{N}}$ are orthonormal, the projector onto the subspace $\mathscr{E}_{0}$  can  then be  written as, 
\beq
\mathcal{P} = \ket{\phi_{1}}\bra{\phi_{1}}  + \ket{\phi_{2}}\bra{\phi_{2}}  + ...+ \ket{\phi_{N}}\bra{\phi_{N}},
\eeq
which satisfies the relations: $\mathcal{P} = \mathcal{P}^{\dagger}, \;  \mathcal{P}^{2} = \mathcal{P}$. Let the supplementary subspace of $\mathscr{E}_{0}$ is $\mathscr{C}_{0}$ and the corresponding projector $Q$ is defined as $\mathcal{Q} = 1 -\mathcal{P}$, which also  obeys the relations: $\mathcal{Q} = \mathcal{Q}^{\dagger}, \;  \mathcal{Q}^{2} = \mathcal{Q}$. Following this, one can  derive the expression for projection of $G(\xi)$  in (\ref{green})  as
\beq
\mathcal{P} \mathcal{G}(\xi) \mathcal{P} = \frac{\mathcal{P}}{\xi-\mathcal{P}H_{0}\mathcal{P} - \mathcal{P} \mathcal{R}(\xi) \mathcal{P}} \; ,
\label{p_op}
\eeq
where the displacement operator $\mathcal{R}(\xi)$ is defined by 
\beq
\mathcal{R}(\xi) = H^{'} + H^{'} \frac{\mathcal{Q}}{\xi-\mathcal{Q}H_{0}\mathcal{Q} -\mathcal{Q}H^{'}\mathcal{Q}} H^{'} .
\eeq

Based on this construction, the authors in the Ref. [\cite{brion2007}] have shown  the effective Hamiltonian for the lambda system  discussed above (Fig. \ref{sim_Rama}). In this case,

\beq
 H_{0} = 
\begin{pmatrix}
-\frac{\Delta}{2} & 0 & 0\\
0 & \frac{\Delta}{2} & 0\\
0 & 0 & \bar{\Delta}
\end{pmatrix}, \quad 
H^{'}= \frac{1}{2} 
\begin{pmatrix}
0 & 0 & \Omega^{\ast}_{1}\\
0 & 0 & \Omega^{\ast}_{2}\\
\Omega_{1} & \Omega_{2} &  0
\end{pmatrix}.
\label{7}
\eeq
Here,  the projectors are read:  $\mathcal{P} =  \ket{g} \bra{g} + \ket{2} \bra{2}$ and $\mathcal{Q}  = \ket{1} \bra{1}$.
 After little algebra, they have shown the effective Hamiltonian $H_{\text{eff}} = \mathcal{ P}H_{0} \mathcal{P} + \mathcal{P} \mathcal{R}(0) \mathcal{P}$, which describes the reduced dynamics of the system in terms of two-photon Raman transition.
\par
Although, this method has been utilized to study an effective operator formalism for open quantum systems [\cite{ISD, FRS}], however, to our knowledge, this approach does not include the cavity mode where the applied laser fields are treated classically in a more general situation. The pole analysis will be cumbersome task when the cavity modes will be incorporated in this situation. As a result, extending this approach to eliminate nonresonant levels in the case of a multiphoton transition in an atom will be a difficult undertaking.

\subsection{Integro-differential equation approach of Paulisch et al}
\label{englart}
In this approach, the authors in the Ref. [\cite{paulisch2014}] have considered a multi-level  system where the atomic transition between the successive levels happens under the application of classical laser fields. They have  used an integro-differential equation approach  for the probability amplitudes for the atomic levels which are resonantly coupled to each other. The integro-differential equation  is then solved  under Markov approximation to adiabatically eliminate the nonresonantly coupled states.  For instance, in case of aforesaid lambda system discussed in subsection \ref{projection},  the effective Hamiltonian  in the unperturbed  basis $\{\ket{g},\ket{2}\}$ is shown to be
\beq
\mathcal{H}_{\text{eff}} = -\dfrac{1}{2}
\begin{pmatrix}
\Delta + \frac{|\Omega_{1}|^{2}}{2\bar{\Delta}} & \frac{\Omega_{1}^{\ast}\Omega_{2}^{\ast}}{2\Delta}\\
\frac{\Omega_{1}\Omega_{2}}{2\Delta} & -\Delta + \frac{|\Omega_{1}|^{2}}{2\bar{\Delta}}
\end{pmatrix}.
\eeq

Despite the fact that this method has been used to investigate the effective dynamics in squeezed-light-enhanced atom interferometry [\cite{SSS}], adiabatic elimination for strong field light matter coupling [\cite{BKT}], third-order diffraction in Raman scattering [\cite{SHJ}], etc., it does not include the cavity modes. In this approach, they have worked in interaction picture and to make the Hamiltonian time independent, a suitable  transformation has been utilized, which will be difficult to obtain when the cavity interactions are present. Furthermore, in the presence of cavity modes, the Schr{\"o}dinger equation for probability amplitudes for the states becomes complicated. However, we demonstrate how to construct an effective Hamiltonian in this general scenario in our technique.

\subsection{ James' and  generalized James' effective Hamiltonian approaches}
\label{james}

In this subsection,  we  discuss  the merits and demerits of the James' and generalized James' effective Hamiltonian methods [\cite{james2, gjames}] briefly. The authors in the Ref.  [\cite{james2}] have considered the time-averaged dynamics of the system to discard the various highly oscillating terms present in the effective Hamiltonian. Keeping the terms up to second order of the interacting Hamiltonian in the time-ordered expansion of the evolution operator, they have shown the effective Hamiltonian as 
\beq
\mathcal{H}_{\text{eff}}^{(2)}(t)\;\; =  \; \;\sum\limits_{\alpha,\beta=1}^{N} \frac{1}{ \bar{\omega}_{_{\alpha \beta}}} \big[\Lambda_{\alpha}^{\dagger},\Lambda_{\beta}\big] \exp(i(\omega_{\alpha} -\omega_{\beta})t),
\label{james_eff}
\eeq
where $\Lambda_{\alpha}$ is the relevant atomic transition operators,  $N$ is the total number of different harmonic term and $\bar{\omega}_{_{\alpha \beta}}$ is harmonic average of $\omega_{\alpha}$ and $\omega_{\beta}$ i.e. $\frac{1}{\bar{\omega}_{_{\alpha \beta}}}\; =\; \frac{1}{2} \left(\frac{1}{\omega_{\alpha}} + \frac{1}{\omega_{\beta}}\right)$.
Although this approach explains two-photon  Raman transition in a three-level system, quantum AC Stark shits [\cite{helom}], single-qubit and two-qubit holonomic gates [\cite{zhao2018}] etc.,  it does not  demonstrate the phenomena related to three-photon or higher order transition processes.

\par
Later on, adopting James' idea, W. Shao et al [\cite{gjames}] have extended the James' effective Hamiltonian method  up to third-order in time-ordered expansion of the evolution operator. By neglecting the rapidly oscillating terms under rotating wave approximation, they have derived the effective Hamiltonian which reads
\bea
\mathcal{H}_{\text{eff}}^{(3)}(t) &=&  \sum\limits_{\alpha, \beta, \gamma} \big\lgroup \frac{1}{\omega_{\gamma} (\omega_{\gamma} -\omega_{\beta})} \big[ \Lambda_{\alpha}\Lambda_{\beta}\Lambda_{\gamma}\exp{(i(\omega_{\alpha}-\omega_{\beta}+\omega_{\gamma})t)} + \Lambda_{\alpha}^{\dagger}\Lambda_{\beta}\Lambda_{\gamma}^{\dagger}\exp{(i(\omega_{\alpha}-\omega_{\beta}+\omega_{\gamma})t)} \nn\\
&+& \Lambda_{\alpha}\Lambda_{\beta}\Lambda_{\gamma}^{\dagger}\exp{(i(\omega_{\alpha} + \omega_{\beta}-\omega_{\gamma})t)} + \Lambda_{\alpha}^{\dagger}\Lambda_{\beta}^{\dagger}\Lambda_{\gamma}\exp{(i(-\omega_{\alpha}-\omega_{\beta}+\omega_{\gamma})t)}\big] + \frac{1}{\omega_{\gamma} (\omega_{\gamma} +\omega_{\beta})}\times \nn\\
&\times & \big[ \Lambda_{\alpha}^{\dagger}\Lambda_{\beta}\Lambda_{\gamma}\exp{(i(-\omega_{\alpha}+\omega_{\beta}+\omega_{\gamma})t)}  + \Lambda_{\alpha}\Lambda_{\beta}^{\dagger}\Lambda_{\gamma}^{\dagger}\exp{(i(\omega_{\alpha}-\omega_{\beta}-\omega_{\gamma})t)} \big]\big\rgroup.
\label{gen_james_eff}
\eea
It is worth noting that to prove the Hermiticity of $\mathcal{H}_{\text{eff}}^{(3)}(t)$  (\ref{gen_james_eff}), the authors in the Ref. [\cite{gjames}]  have introduced the condition that all of the frequencies $\omega_{m}$ are distinct, and the algebraic sum of any three frequencies is zero i.e.
\beq
\omega_{\alpha} +\omega_{\beta}- \omega_{\gamma} = 0.
\label{gen_james_con}
\eeq 
We observe that this condition of Hermiticity  actually leads to the transitions at resonance which we shall  demonstrate in Sec. \ref{3photon} in detail. Because of this condition of Hermiticity, the generalized James' effective Hamiltonian method does not  explain  off-resonant transitions. However, we propose an approach based on the  adiabatic elimination under Markov approximation which includes the transitions  both at off-resonance and as well as at resonance.

\section{ Adiabatic elimination in the Heisenberg picture using Markov approximation}
\label{Markov_approx}
As we have pointed out in the previous sections, the adiabatic elimination techniques already existing in the literature work very well in the case of few level systems interacting with classical laser fields. However, these are either not suitable or are too cumbersome to deal with multiphoton resonances in multilevel systems in the presence of interactions with cavity modes. 
\par
We propose here an alternative approach to the problem of deriving effective Hamiltonians in the case of atomic systems interacting with both the classical laser fields as well as quantized cavity fields, by adiabatically eliminating nonresonantly coupled n-photon intermediate transitions. We use an adaptation of the Markovian approach as discussed earlier by Paulisch et al [\cite{paulisch2014}]. In contrast to the previous approaches, we work entirely in the Heisenberg picture. While the cavity modes are represented by the corresponding creation and annihilation operators ($a^{\dagger}$,  $a$), the atomic systems are represented by appropriate transition operators such as $\sigma_{\imath \jmath} \equiv \ket{\imath} \bra{\jmath}$, where $\ket{\imath}$ and $\ket{\jmath}$ are some unperturbed energy states. For the sake of illustration, let us consider the simple case of a transition from level $\ket{g}$ to level $\ket{N-1}$ by absorption of photons from laser fields with frequencies, $\omega_{\jmath}  (\jmath = 1,2,....,N-1)$. The levels $\ket{N-1}$ and $\ket{N}$ are connected by a one-photon transition induced by single cavity mode of frequency $\omega_{c}$ and Rabi frequency $\eta $. The Hamiltonian describing this system is given by 

\begin{figure}[hbt!]
	\begin{center}
	\captionsetup[subfigure]{labelformat=empty}
	\includegraphics[width=7.5cm,height=6cm]{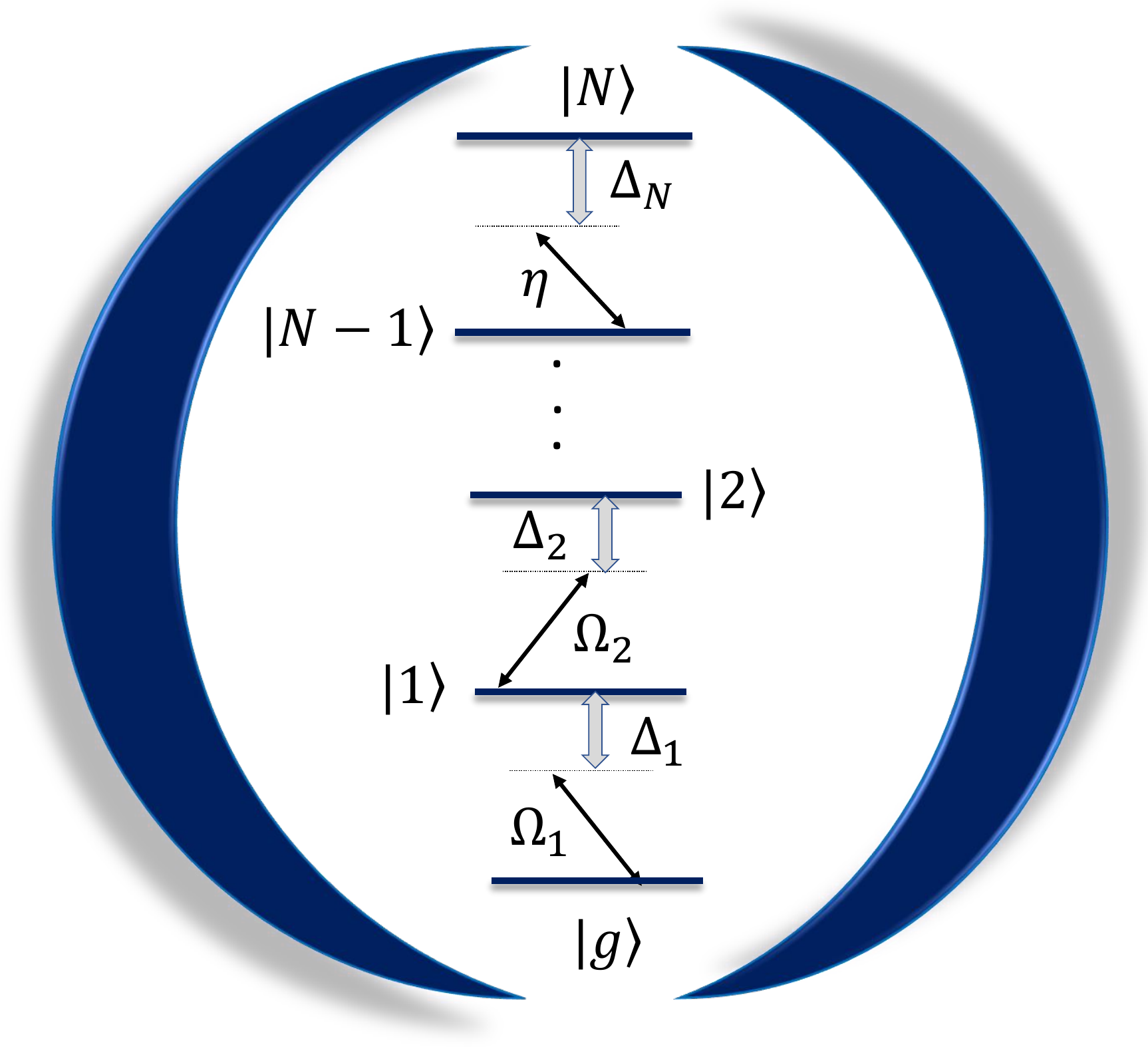} 
	\caption{ Energy-level diagram for n-photon transitions in atoms inside a microwave cavity under the application of classical laser pulses and the cavity mode.} 
\label{nphoton_markov}
\end{center}
\end{figure}

\bea
H  &=& H_{0} + H^{'},
\label{Hami1}
\eea
where
\bea
H_{0} &=& \sum\limits_{\jmath=1}^{N} E_{\jmath} \sigma_{\jmath \jmath} + E_{g} \sigma_{gg} + \omega_{c} a^{\dagger} a, 
\label{Hami11}
\eea
is the free atom plus free field Hamiltonian and 
\bea
H^{'} &=& \sum\limits _{\jmath = 1}^{ N-2} \big\{\Omega_{\jmath+1}(t) \exp{(-i \omega_{\jmath+1}t)} \sigma_{\jmath+1, \jmath} +  \Omega_{\jmath+1}^{\ast}(t) \exp{(i \omega_{\jmath+1}t)} \sigma_{\jmath, \jmath+1}\big\} + \big(\Omega_{_{1}} (t) \exp{(-i\omega_{1} t)} \sigma_{_{1g}}  \nn \\
&+& \Omega_{_{1}}^{\ast} (t) \exp{(i\omega_{_{1}} t)} \sigma_{_{g1}}\big) + \left( \eta  a \sigma_{_{N, N-1}} + \eta  a^{\dagger} \sigma_{_{N-1, N}}\right)
\label{Hami12}
\eea 
is the interaction Hamiltonian. For simplicity, we assume the energy of the level $\ket{g}$, $E_{g}  = 0$, which implies $E_{\jmath}  = \hbar \omega_{\jmath g}$ and $E_{_{N}} = \hbar\omega_ {_{Ng}}$. We further assume that $E_{\jmath} < E_{\jmath+1}$ and $E_{_{N-1}} < E_{_{N}}$.  In the interaction picture defined with  $H_{0}$ (\ref{Hami11}), the reduced form of the Hamiltonian (\ref{Hami1}) can be written as ($\hbar =1$\; herein)
\bea
H_{_{I}} &=& \sum\limits _{\jmath = 1}^{ N-2} \big\{\Omega_{\jmath+1}(t) \exp{(i \Delta_{\jmath+1}t)} \sigma_{\jmath+1, \jmath} +  \Omega_{\jmath+1}^{\ast}(t) \exp{(-i \Delta_{\jmath+1}t)} \sigma_{\jmath, \jmath+1}\big\} + \big(\Omega_{1} (t) \exp{(i\Delta_{1} t)} \sigma_{_{1g}}  \nn \\
&+& \Omega_{1}^{\ast} (t) \exp{(-i\Delta_{1} t)} \sigma_{_{g1}}\big) + \left( \eta \exp{(i \Delta_{_{N}} t)} a \sigma_{_{N, N-1}} + \eta \exp{( - i \Delta_{_{N}}t)}  a^{\dagger} \sigma_{_{N-1, N}}\right),
\label{Hami_inter}
\eea 
where the detunings, $\Delta_{\jmath+1}  = \left(\omega_{\jmath +1, g} - \omega_{\jmath g}\right) - \omega_{\jmath + 1}$; $\Delta_{1}  = \omega_{1g} - \omega_{1}$ and $\Delta_{_{N}}  = \omega_{_{Ng}} - \omega_{c} $.
\par
Using the Hamiltonian (\ref{Hami_inter}), we obtain the Heisenberg equations for the operators $ \sigma_{\jmath,\jmath+1}$, $\sigma_{_{g1}}$ and $\sigma_{_{N-1, N}}$
\bea
i\dot{\sigma}_{{\jmath,\jmath+1}} &=& \sum\limits _{\imath = 1}^{ N-2}   \big\{\Omega_{\imath+1}(t) \exp{(i \Delta_{\imath+1}t)} \left( \sigma_{\jmath \imath} \delta_{\jmath +1,\imath +1} - \sigma_{\imath +1,\jmath +1} \delta_{\imath \jmath}\right) +   \Omega_{\imath+1}^{\ast}(t) \exp{(-i \Delta_{\imath+1}t)} \qquad \quad \nn\\\
&&\times \left( \sigma_{\jmath, \imath+1} \delta_{\jmath +1,\imath} - \sigma_{\imath,\jmath +1} \delta_{\imath+1, \jmath}\right)\big\} -\Omega_{1}^{\ast}(t) \exp{(-i \Delta_{1} t)} \sigma_{g,\jmath+1} \delta_{1\jmath} \qquad \quad \nn\\
&& +\;\; \eta \exp{(-i \Delta_{_{N}} t)} a^{\dagger} \sigma_{_{\jmath N}} \delta_{\jmath+1, N-1},
\label{s11}
\eea
\bea
i\dot{\sigma}_{_{g1}} & =&  \sum\limits _{\jmath = 1}^{ N-2}  \Omega_{\jmath+1}^{\ast}(t) \exp{(-i \Delta_{\jmath+1}t)}\sigma_{g, \jmath+1} \delta_{1\jmath} +  \Omega_{1}(t) \exp{(i \Delta_{1} t)} (\sigma_{gg} - \sigma_{11}), \qquad \qquad \qquad \;\;\;
\label{s12}
\eea
\bea
i\dot{\sigma}_{_{N-1, N}} & =&  \sum\limits _{\jmath = 1}^{ N-2}  - \; \Omega_{\jmath+1}^{\ast}(t) \exp{(-i \Delta_{\jmath+1}t)}\sigma_{\jmath N} \delta_{\jmath+1,N-1} +  \eta \exp{(-i \Delta_{_{N}} t)} a(\sigma_{_{N-1, N-1}} - \sigma_{_{NN}}). \qquad  \;\;\;\;
\label{s13}
\eea
Now further assuming the detunings  $\Delta_{\jmath}\; (\jmath = 1,2,...,N) $ to be large compare to $|\Omega_{\jmath}(t)|$ and $\eta $, levels $ \ket{1}, \ket{2}, .....,\ket{N-1} $ may be adiabatically eliminated. To this end, for solving the Eqs. (\ref{s11}-\ref{s13}), we make use of  Markov approximation which allows us to consider $\Omega_{\jmath}(t^{'}) \approx \Omega_{\jmath}(t)$. Therefore from Eq.(\ref{s11}), we have
\bea
\sigma_{\jmath,\jmath+1} &=& \sum\limits _{\imath = 1}^{ N-2}   \big\{-\frac{\Omega_{\imath+1}(t)} {\Delta_{\imath+1}}\exp{(i \Delta_{\imath+1}t)} \left( \sigma_{\jmath \imath} \delta_{\jmath +1,\imath +1} - \sigma_{\imath +1,\jmath +1} \delta_{\imath \jmath}\right) +   \frac{\Omega_{\imath+1}^{\ast}(t)}{\Delta_{\imath+1}} \exp{(-i \Delta_{\imath+1}t)}\nn\\\!\!\!\!\!\!\!\!
&&\times \; \left( \sigma_{\jmath, \imath+1} \delta_{\jmath +1,\imath} - \sigma_{\imath,\jmath +1} \delta_{\imath+1, \jmath}\right)\big\} -\frac{\Omega_{1}^{\ast}(t)}{\Delta_{1}} \exp{(-i \Delta_{1} t)} \sigma_{g,\jmath+1} \delta_{1\jmath} \nn\\
&&+\;  \frac{\eta}{\Delta_{_{N}}} \exp{(-i \Delta_{_{N}} t)} a^{\dagger} \sigma_{_{\jmath N}} \delta_{\jmath+1, N-1},
\label{sol_s11}
\eea
Following the above approximation, from the Eqs.(\ref{s12}) and (\ref{s13}) we obtain,
\bea
\sigma_{_{g1}} & =&  \sum\limits _{\jmath = 1}^{ N-2} \; \frac{\Omega_{\jmath+1}^{\ast}(t)}{\Delta_{\jmath+1}} \exp{(-i \Delta_{\jmath+1}t)}\sigma_{g, \jmath+1} \delta_{1\jmath} -  \frac{\Omega_{1}(t)}{\Delta_{1}} \exp{(i \Delta_{1} t)} (\sigma_{gg} - \sigma_{11}), \qquad \qquad \qquad \;\;\;
\label{sol_s12}
\eea
\bea
\sigma_{_{N-1, N}} & =&  \sum\limits _{\jmath = 1}^{ N-2}  -\; \frac{\Omega_{\jmath+1}^{\ast}(t)}{\Delta_{\jmath+1}} \exp{(-i \Delta_{\jmath+1}t)}\sigma_{_{\jmath N}} \delta_{\jmath+1, N-1} +  \frac{\eta}{\Delta_{_{N}}}\exp{(-i \Delta_{_{N}} t)} a (\sigma_{_{N-1,N-1}} - \sigma_{_{NN}}), \quad \quad \;
\label{sol_s13}
\eea
respectively. Now assuming that Stark shifts resulting from the diagonal terms in (\ref{sol_s11}-\ref{sol_s13}) are small we have
\bea
\sigma_{\jmath,\jmath+1} &\simeq& \sum\limits _{\imath = 1}^{ N-2} \bigg\{\frac{\Omega_{\imath+1}^{\ast}(t)}{\Delta_{\imath+1}} \exp{(-i \Delta_{\imath+1}t)} \left( \sigma_{\jmath, \imath+1} \delta_{\jmath +1,\imath} - \sigma_{\imath,\jmath +1} \delta_{\imath+1, \jmath}\right)\bigg\} \nn\\
&-&\frac{\Omega_{1}^{\ast}(t)}{\Delta_{1}} \exp{(-i \Delta_{1} t)} \sigma_{g,\jmath+1} \delta_{1\jmath} + \frac{\eta}{\Delta_{_{N}}} \exp{(-i \Delta_{_{N}} t)} a^{\dagger} \sigma_{_{\jmath N}} \delta_{\jmath+1, N-1}, \qquad \qquad \qquad \qquad
\label{re_sol_s11}
\eea
\bea
\sigma_{_{g1}}  &\simeq &\frac{\Omega_{2}^{\ast}(t)}{\Delta_{2}} \exp{(-i \Delta_{2}t)}\sigma_{_{g 2}}, \qquad \qquad \qquad \qquad \qquad\qquad \qquad \qquad \qquad \qquad \qquad \quad \;\;\;
\label{re_sol_s12}
\eea
\bea
\sigma_{_{N-1, N}} & \simeq& - \frac{\Omega_{N}^{\ast}(t)}{\Delta_{N}} \exp{(-i \Delta_{N}t)}\sigma_{_{N-1, N}}. \qquad \qquad \qquad \qquad \qquad\qquad \qquad \qquad \qquad \qquad \qquad\;\;\;
\label{re_sol_s13}
\eea
\par 
Inserting these results into the equation (\ref{Hami_inter}), we have
\bea
H_{_{I}}^{(1)}  &= & \sum\limits _{\jmath = 1}^{ N-1} \big\{\Omega_{\jmath+1}^{\ast}(t) \Omega_{\jmath+2}^{\ast}(t) \left( \frac{1}{\Delta_{\jmath+2}} -\frac{1}{\Delta_{\jmath+1}}\right) \exp{\left(-i( \Delta_{\jmath+1} + \Delta_{\jmath+2})t\right)} \sigma_{\jmath, \jmath+2}  + H.c. \big\}\nn\\
&+& \Omega_{1}^{\ast}(t) \Omega_{2}^{\ast}(t) \left(\frac{1}{\Delta_{2}} -\frac{1}{\Delta_{1}}\right) \exp{\left(-i(\Delta_{1}+\Delta_{2}) t\right)} \sigma_{g2}  \nn\\
&+& \eta\; \Omega_{_{N-1}}^{\ast}(t) \left(\frac{1}{\Delta_{_{N}}} -\frac{1}{\Delta_{_{N-1}}}\right) \exp{\left(-i( \Delta_{_{N-1}}+\Delta_{_{N}}) t\right)}\; a^{\dagger} \sigma_{_{N-2,N}} + H.c.
\label{m1_Hami_inter}
\eea
At this point the Hamiltonian has three parts, the first and second  parts correspond to the effective two-photon transitions between  $\ket{\jmath} \leftrightarrow \ket{\jmath +2} $ and $\ket{g } \leftrightarrow \ket{2}$ respectively while the third  one gives the usual Raman transitions between transitions between the states $ \ket{N-1}$ and $\ket{N}$. To pursue the adiabatic elimination of the intermediate levels, we further need the equations for $\sigma_{\jmath,\jmath+2}$, $\sigma_{g2}$ and $\sigma_{_{N-1,N}}$. 
\par
Following the above procedure we find 
\bea
\sigma_{\jmath,\jmath+2} &=& \sum\limits _{\imath = 1}^{ N-2}   \bigg\{-\frac{\Omega_{\imath+1}(t)} {\Delta_{\imath+1}}\exp{(i \Delta_{\imath+1}t)} \left( \sigma_{\jmath \imath} \delta_{\jmath +2,\imath +1} - \sigma_{\imath +1,\jmath +2}\; \delta_{\imath \jmath}\right) +   \frac{\Omega_{\imath+1}^{\ast}(t)}{\Delta_{\imath+1}} \exp{(-i \Delta_{\imath+1}t)}\nn\\\!\!\!\!\!\!\!\!
&&\times \; \left( \sigma_{\jmath, \imath+2}\; \delta_{\jmath +2,\imath} - \sigma_{\imath,\jmath +2} \;\delta_{\imath+1, \jmath}\right)\bigg\} -\frac{\Omega_{1}^{\ast}(t)}{\Delta_{1}} \exp{(-i \Delta_{1} t)} \sigma_{g,\jmath+2} \;\delta_{1\jmath} \nn\\
&&+\;  \frac{\eta}{\Delta_{_{N}}} \exp{(-i \Delta_{_{N}} t)}\; a^{\dagger} \sigma_{_{\jmath N}} \delta_{\jmath+2, N-1},
\label{sol_s21}
\eea
\bea
\sigma_{_{g2}} &=&  \sum\limits _{\jmath = 1}^{ N-2} -\frac{\Omega_{\jmath+1}(t)}{\Delta_{\jmath+1}} \exp{(i \Delta_{\jmath+1}t)}\sigma_{g \jmath} \;\delta_{\jmath+1,2} +  \frac{\Omega_{\jmath+1}^{\ast}(t)}{\Delta_{\jmath+1}} \exp{(i \Delta_{\jmath+1}t)}\sigma_{g, \jmath+1} \delta_{2\jmath} \qquad \qquad \;\;\;\; \nn\\ 
 &+ & \frac{\Omega_{1}(t)}{\Delta_{1}} \exp(i \Delta_{1}t) \sigma_{_{12}}, 
\label{sol_s22}
\eea
\bea
\sigma_{_{N-2,N}} &=& \sum\limits _{\jmath = 1}^{ N-2}   \bigg\{-\;\frac{\Omega_{\jmath+1}(t)} {\Delta_{\jmath+1}}\exp{(i \Delta_{\jmath+1}t)} \left( \sigma_{_{N-2,\jmath}} \; \delta_{N,\jmath +1} - \sigma_{\jmath +1, N} \; \delta_{\jmath, N-2}\right)\qquad \qquad\qquad \qquad \quad\nn \\ &&+\;\frac{\Omega_{\jmath+1}^{\ast}(t)}{\Delta_{\jmath+1}} \exp{(-i \Delta_{\jmath+1}t)}\left( \sigma_{_{N-2, \jmath+1}}\; \delta_{N \jmath}- \sigma_{\jmath N}\; \delta_{\jmath+1, N-2}\right) \bigg\} \nn\\
&&+ \;\frac{\eta}{\Delta_{_{N}}} \exp{(-i \Delta_{_{N}}t)}\; a^{\dagger} \sigma_{_{N-2, N-1}}.\qquad \qquad\qquad \qquad \qquad\qquad
\label{sol_s23}
\eea
As we have already  assumed that the detunings $\Delta_{\jmath}$ are large, therefore neglecting the terms on the order of $(\Delta_{\imath} \Delta_{\jmath})^{-1} \; (\imath, \jmath  = 1, 2, ...., N)$  in (\ref{sol_s21} -\ref{sol_s23}) we have
\bea
\sigma_{\jmath,\jmath+2} &\simeq & \sum\limits _{\imath = 1}^{ N-1} \bigg\{\frac{\Omega_{\imath+1}^{\ast}(t)}{\Delta_{\imath+1}} \exp{(-i \Delta_{\imath+1}t)} \left( \sigma_{\jmath, \imath+1} \delta_{\jmath +2,\imath} - \sigma_{\imath,\jmath +2} \delta_{\imath+1, \jmath}\right)\bigg\} \nn\\
&-&\frac{\Omega_{1}^{\ast}(t)}{\Delta_{1}} \exp{(-i \Delta_{1} t)} \sigma_{g,\jmath+2}\; \delta_{1\jmath} + \frac{\eta}{\Delta_{_{N}}} \exp{(-i \Delta_{_{N}} t)}\; a^{\dagger} \sigma_{_{\jmath N}} \delta_{\jmath+2, N-1} , \qquad \qquad \qquad \qquad \qquad\qquad\;\
\label{re_sol_s21}
\eea
\bea
\sigma_{_{g2}}  &\simeq &\frac{\Omega_{3}^{\ast}(t)}{\Delta_{3}} \exp{(-i \Delta_{3}t)}\sigma_{_{g 3}}, \qquad \qquad \qquad \qquad \qquad\qquad \qquad \qquad \qquad \qquad \qquad \quad \;\;\;
\label{re_sol_s22}
\eea
\bea
\sigma_{_{N-2,N}} & \simeq & - \frac{\Omega_{N-2}^{\ast}(t)}{\Delta_{_{N-2}}} \exp{(-i \Delta_{_{N-2}}t)}\sigma_{_{N-3, N}}. \qquad \qquad \qquad \qquad \qquad\qquad \qquad \qquad \qquad \quad\;\;\;\;
\label{re_sol_s23}
\eea
Inserting these results into the Eq. (\ref{m1_Hami_inter}), we obtain 
\bea
H_{_{I}}^{(2)} & =&  \sum\limits _{\jmath = 1}^{ N-2} \bigg\{ \Omega_{\jmath+1}^{\ast}(t) \Omega_{\jmath+2}^{\ast}(t) \Omega_{\jmath+3}^{\ast}(t) \left( \frac{1}{\Delta_{\jmath+3}} \left( \frac{1}{\Delta_{\jmath+2}} -\frac{1}{\Delta_{\jmath+1}}\right)- \frac{1}{\Delta_{\jmath+1}} \left( \frac{1}{\Delta_{\jmath+3}} -\frac{1}{\Delta_{\jmath+2}}\right)\right) \nn\\
&\times & \exp{\left(-i \left(\Delta_{\jmath+1} + \Delta_{\jmath+2}+\Delta_{\jmath+3} \right) t\right)} \sigma_{\jmath,\jmath+3} + \text{H.c.} \bigg\} \nn\\
&+& \Omega_{1}^{\ast}(t) \Omega_{2}^{\ast}(t) \Omega_{3}^{\ast}(t) \left( \frac{1}{\Delta_{3}} \left( \frac{1}{\Delta_{2}} -\frac{1}{\Delta_{1}}\right)- \frac{1}{\Delta_{1}} \left( \frac{1}{\Delta_{3}} -\frac{1}{\Delta_{2}}\right)\right)
\times \exp{\left(-i \left(\Delta_{1} + \Delta_{2}+\Delta_{3} \right) t\right)} \sigma_{_{g3}} \nn\\
& + & \eta \; \Omega_{N-1}^{\ast}(t) \Omega_{N-2}^{\ast}(t) \left( \frac{1}{\Delta_{_{N}}} \left( \frac{1}{\Delta_{_{N-1}}} - \frac{1}{\Delta_{_{N-2}}}\right)- \frac{1}{\Delta_{_{N-2}}} \left( \frac{1}{\Delta_{_{N}}} -\frac{1}{\Delta_{_{N-1}}}\right)\right)\nn\\
&\times& \exp{\left(-i \left(\Delta_{_{N-2}} + \Delta_{_{N-1}}+ \Delta_{_{N}} \right) t\right)} \;  a^{\dagger}\sigma_{_{N-3,N}} + \text{H.c.} 
\label{m2_Hami_inter}
\eea
As above, the first two terms in the Hamiltonian represent an effective three-photon transitions between \; $\ket{\jmath} \leftrightarrow \ket{\jmath+3}$  and $\ket{g} \leftrightarrow \ket{3}$ while the last one describes the three-photon hyper Raman transitions between $\ket{N-3}$ and $\ket{N}$. From this point, it is clear that to eliminate next levels we need to calculate the equations of motion for the operators  $\sigma_{\jmath,\jmath+3}$, $\sigma_{_{g3}}$ and $\sigma_{_{N-3,N}}$. 
\par  
Therefore, following the above procedure and based on  the assumption adopted in  case of Eqs. (\ref{re_sol_s21}-\ref{re_sol_s23}), one can obtain the approximate solutions  for  $\sigma_{\jmath,\jmath+3}$, $\sigma_{g3}$ and $\sigma_{_{N-3,N}}$. Inserting these results into (\ref{m2_Hami_inter}) we have
\bea
H_{_{I}}^{(3)}\;\; =\;\;  \sum\limits _{\jmath = 1}^{ N-2} \bigg\{ \bar{\Omega}_{\jmath}\Bigg(\frac{1}{\Delta_{\jmath+4}}\left( \frac{1}{\Delta_{\jmath+3}} \left( \frac{1}{\Delta_{\jmath+2}} -\frac{1}{\Delta_{\jmath+1}}\right)- \frac{1}{\Delta_{\jmath+1}} \left( \frac{1}{\Delta_{\jmath+3}} -\frac{1}{\Delta_{\jmath+2}}\right)\right) - \frac{1}{\Delta_{\jmath+1}} \qquad \qquad\;\;\;\nn \\ 
\times\left( \frac{1}{\Delta_{\jmath+4}} \left( \frac{1}{\Delta_{\jmath+3}} -\frac{1}{\Delta_{\jmath+2}}\right)- \frac{1}{\Delta_{\jmath+2}} \left( \frac{1}{\Delta_{\jmath+4}} -\frac{1}{\Delta_{\jmath+3}}\right)\right)\Bigg) \exp{\left(-i \bar{\Delta}_{\jmath} t\right)}\; \sigma_{\jmath,\jmath+4} + \text{H.c}\Bigg\}\;\;\;\;\nn\\
+\;\; \bar{\Omega}_{0} \Bigg(\frac{1}{\Delta_{4}}\left( \frac{1}{\Delta_{3}} \left( \frac{1}{\Delta_{2}} -\frac{1}{\Delta_{1}}\right)- \frac{1}{\Delta_{1}} \left( \frac{1}{\Delta_{3}} -\frac{1}{\Delta_{2}}\right)\right) - \frac{1}{\Delta_{1}} \Bigg( \frac{1}{\Delta_{4}} \left( \frac{1}{\Delta_{3}} -\frac{1}{\Delta_{2}}\right)- \qquad \qquad \;\;\nn\\
- \frac{1}{\Delta_{2}} \left( \frac{1}{\Delta_{4}} -\frac{1}{\Delta_{3}}\right)\Bigg)\Bigg) \exp{\left(-i \bar{\Delta}_{0} t\right)}\; \sigma_{_{g4}} + \text{H.c.} \qquad \qquad\qquad \qquad\qquad \qquad\qquad\;\;\;\;\;\;\; \nn\\
+\; \bar{\eta}\;
\Bigg(\frac{1}{\Delta_{_{N}}}\Bigg(\frac{1}{\Delta_{_{N-1}}}\left(\frac{1}{\Delta_{_{N-2}}}+\frac{1}{\Delta_{_{N-3}}}\right) - \frac{1}{\Delta_{_{N-3}}}\left(\frac{1}{\Delta_{_{N-1}}}-\frac{1}{\Delta_{_{N-2}}}\right)\Bigg) - \frac{1}{\Delta_{_{N-3}}}\times \quad \quad \quad\qquad \nn \\
\times \;\Bigg(\frac{1}{\Delta_{_{N}}}\left( \frac{1}{\Delta_{_{N-1}}}-\frac{1}{\Delta_{_{N-2}}}\right) 
- \frac{1}{\Delta_{_{N-2}}}\left( \frac{1}{\Delta_{_{N}}}-\frac{1}{\Delta_{_{N-1}}}\right)\Bigg)\Bigg) \exp{\left(\ -i  \bar{\Delta}_{_{N-4}}t\right)}\; a^{\dagger}  \sigma_{_{N-4,N}} + \text{H.c.},
\label{m3_Hami_inter}
\eea
where the abbreviations:  $ \bar{\Omega}_{\jmath}  = \Omega_{\jmath+1}^{\ast}(t) \Omega_{\jmath+2}^{\ast}(t) \Omega_{\jmath+3}^{\ast}(t)\Omega_{\jmath+4}^{\ast}(t)$, \; $\bar{\Delta}_{\jmath}  =\Delta_{\jmath+1} + \Delta_{\jmath+2}+\Delta_{\jmath+3} + \Delta_{\jmath+4 }$ and $\bar{\eta}  = \eta\; \Omega_{N-3}^{\ast}(t)\;\Omega_{N-2}^{\ast}(t)\;\Omega_{N-1}^{\ast}(t) $. Now, by observing the Eqs. (\ref{m1_Hami_inter}), (\ref{m2_Hami_inter}) and (\ref{m3_Hami_inter}), one can find that there exists  recurrence relations between the coefficients of successive transition operators. For instance, let us first consider the coefficients of the transition operators  $\sigma_{\jmath,\jmath+2}$, $\sigma_{\jmath,\jmath+3}$ and $\sigma_{\jmath,\jmath+4}$. Let's define $\tilde{\Omega}_{\jmath}^{(1)}  = \Omega_{\jmath+1}^{\ast}(t) \Omega_{\jmath+2}^{\ast}(t)$ and  $\tilde{\Delta}_{\jmath}^{(1)} = \Delta_{\jmath+1}+\Delta_{\jmath+2}$ in the Eq. (\ref{m1_Hami_inter}). With these, from the Eqs.(\ref{m2_Hami_inter}) and (\ref{m3_Hami_inter}), we have $\tilde{\Omega}_{\jmath}^{(2)}  = \Omega_{\jmath+3}^{\ast}(t) \tilde{\Omega}_{\jmath}^{(1)}$, $\tilde{\Delta}_{\jmath}^{(2)} = \Delta_{\jmath+3} + \tilde{\Delta}_{\jmath}^{(1)}$ and $\tilde{\Omega}_{\jmath}^{(3)}  = \Omega_{\jmath+4}^{\ast}(t) \tilde{\Omega}_{\jmath}^{(2)}$, $\tilde{\Delta}_{\jmath}^{(3)} = \Delta_{\jmath+4} + \tilde{\Delta}_{\jmath}^{(2)}$ respectively.  Moreover, let us define $\mathcal{C}_{\jmath}^{(1)}  = \frac{1}{\Delta_{\jmath+2}} -\frac{1}{\Delta_{\jmath+1}}$ in the Eq. (\ref{m1_Hami_inter}). Thus,  $\mathcal{C}_{\jmath}^{(2)}  = \frac{1}{\Delta_{\jmath+3}} \mathcal{C}_{\jmath}^{(1)} - \frac{1}{\Delta_{\jmath+1}} \mathcal{C}_{\jmath+1}^{(1)} $ and $\mathcal{C}_{\jmath}^{(3)}  = \frac{1}{\Delta_{\jmath+4}} \mathcal{C}_{\jmath}^{(2)} - \frac{1}{\Delta_{\jmath+1}} \mathcal{C}_{\jmath+1}^{(2)} $. It is noted that $\jmath = 0$  gives  the coefficients of $\sigma_{g2}$, $\sigma_{g3}$ and $\sigma_{g4}$. 
\par  
Furthermore, by looking at the last term in the Eqs. (\ref{m1_Hami_inter}), (\ref{m2_Hami_inter}) and (\ref{m3_Hami_inter}), recurrence relations between the coefficients of successive transition operators can be found. To this end, let us define, $\tilde{\eta}_{_{N}}^{(1)}  =  \eta \; \Omega_{N-1}^{\ast}(t)$  and   using the above definition we have $\mathcal{C}_{_{N-2}}^{(1)}  = \frac{1}{\Delta_{_{N}}} -\frac{1}{\Delta_{_{N-1}}}$ and  $\tilde{\Delta}_{_{N-2}}^{(1)} = \Delta_{_{N}} +\Delta_{_{N-1}} $  in the Eq. (\ref{m1_Hami_inter}). Thus,  from the Eqs. (\ref{m2_Hami_inter}) and (\ref{m3_Hami_inter}), we have $\tilde{\eta}_{_{N}}^{(2)}  =   \Omega_{N-2}^{\ast}(t) \tilde{\eta}_{_{N}}^{(1)}$,  $\tilde{\Delta}_{_{N-2}}^{(2)} = \Delta_{_{N-2}} + \tilde{\Delta}_{_{N-2}}^{(1)}$  and  $\tilde{\eta}_{_{N}}^{(3)}  =   \Omega_{N-3}^{\ast}(t) \tilde{\eta}_{_{N}}^{(2)}$,  $\tilde{\Delta}_{_{N-2}}^{(3)} = \Delta_{_{N-3}} + \tilde{\Delta}_{_{N-2}}^{(2)}$  respectively. Therefore, after performing  $m$ iteration $(m=2, 3,..., N-2)$ of substitution of the solutions of transition operators obtained under Markov approximation to the Hamiltonian in the $(m-1)$ step, one can obtain
\bea
H_{_{I}}^{(m)} & =&  \sum\limits _{\jmath = 1}^{ N-2} \Big\{\tilde{\Omega}_{\jmath}^{(m)} \mathcal{C}_{\jmath}^{(m)} \exp{(-i \tilde{\Delta}_{\jmath}^{(m)} t)} \sigma_{\jmath, \jmath+ m+ 1}  + \text{H.c.}\Big \} + \Big( \tilde{\Omega}_{0}^{(m)} \mathcal{C}_{0}^{(m)} \exp{(-i \tilde{\Delta}_{0}^{(m)} t)} \sigma_{g, m+ 1} \nn \\
&+ &\tilde{\eta}_{_{N}}^{(m)} \mathcal{C}_{_{N-2}}^{(m)} \exp{(-i \tilde{\Delta}_{_{N-2}}^{(m)}t )}\; a^{\dagger} \sigma_{_{N-m-1, N}} + \text{H.c.}\Big), 
\label{m_level_Hami}
\eea
where  \;  $\tilde{\Omega}_{\jmath}^{(m)}  = \Omega_{\jmath+m+1}^{\ast}(t) \tilde{\Omega}_{\jmath}^{(m-1)}$,\;  $\tilde{\Delta}_{\jmath}^{(m)} = \Delta_{\jmath+m+1} + \tilde{\Delta}_{\jmath}^{(m-1)}$, \;$\mathcal{C}_{\jmath}^{(m)}  = \frac{1}{\Delta_{\jmath+m+1}} \mathcal{C}_{\jmath}^{(m-1)} - \frac{1}{\Delta_{\jmath+1}} \mathcal{C}_{\jmath+1}^{(m-1)}$,  \; $\tilde{\eta}_{_{N}}^{(m)} = \tilde{\Omega}_{N-m}^{\ast}(t) \tilde{\eta}_{_{N}}^{(m-1)}$, \; $\tilde{\Delta}_{_{N-2}}^{(m)}  = \Delta_{_{N-m}} + \tilde{\Delta}_{_{N-2}}^{(m-1)}$, and \; $\mathcal{C}_{_{N-2}}^{(m)} = \frac{1}{\Delta_{_{N}}}\mathcal{C}_{_{N-3}}^{(m-1)} -\frac{1}{\Delta_{_{N-m}}} \mathcal{C}_{_{N-2}}^{(m-1)}$. 
Since $\jmath $ runs from $1$ to $N-2$, the first term in (\ref{m_level_Hami}) is ruled out when  $m = N-2$.  Therefore, we have
\bea
H_{_{I}}^{(N-2)} & =&   \tilde{\Omega}_{0}^{(N-2)} \mathcal{C}_{0}^{(N-2)} \exp{(-i \tilde{\Delta}_{0}^{(N-2)} t)} \sigma_{_{g, N-1}} + \tilde{\eta}_{_{N}}^{(N-2)} \mathcal{C}_{_{N-2}}^{(N-2)} \exp{(-i \tilde{\Delta}_{_{N-2}}^{(N-2)}t )}\; a^{\dagger} \sigma_{_{1 N}} + \text{H.c.}.
\label{re_m_level_Hami}
\eea
At this stage, the first term in the Hamiltonian (\ref{re_m_level_Hami}) represents the transition between the levels $g$ and $N-1$ while the second term corresponds to the transition between the levels $1$ and $N$. Finally, using (\ref{Hami_inter}), we calculate the equations of motion for $ \sigma_{_{g, N-1}}$ and $ \sigma_{_{1N}}$ and substituting their solutions into (\ref{re_m_level_Hami}), we obtain the effective Hamiltonian 

\bea
H_{\text{eff}} (t)& =& \eta\; \tilde{\Omega}_{0}^{(N-2)}\left( \frac{1}{\Delta_{_{N}}} \mathcal{C}_{0}^{N-2} - \frac{1}{\Delta_{_{1}}} \mathcal{C}_{_{N-2}}^{(N-2)}\right)\;\exp{\left( -i \tilde{\Delta}_{0}^{(N-1)} t\right)} \; a^{\dagger} \sigma_{_{gN}} + \text{H.c.} \nn\\
&=& \eta\; \tilde{\Omega}_{0}^{(N-2)} \mathcal{C}_{0}^{(N-1)} \exp{\left( -i \tilde{\Delta}_{0}^{(N-1)} t\right)} \; a^{\dagger} \sigma_{_{gN}} + \text{H.c.},
\label{main_effec}
\eea
which describes the effective dynamics between the states  $\ket{g}$ and $\ket{N}$.

\subsection{Revisit the two-photon transition in a three-level atom inside a cavity using our approach} 
\label{two_photon}
\begin{figure}[hbt!]
	\begin{center}
	\captionsetup[subfigure]{labelformat=empty}
	\includegraphics[width=6.8cm,height=5.0cm]{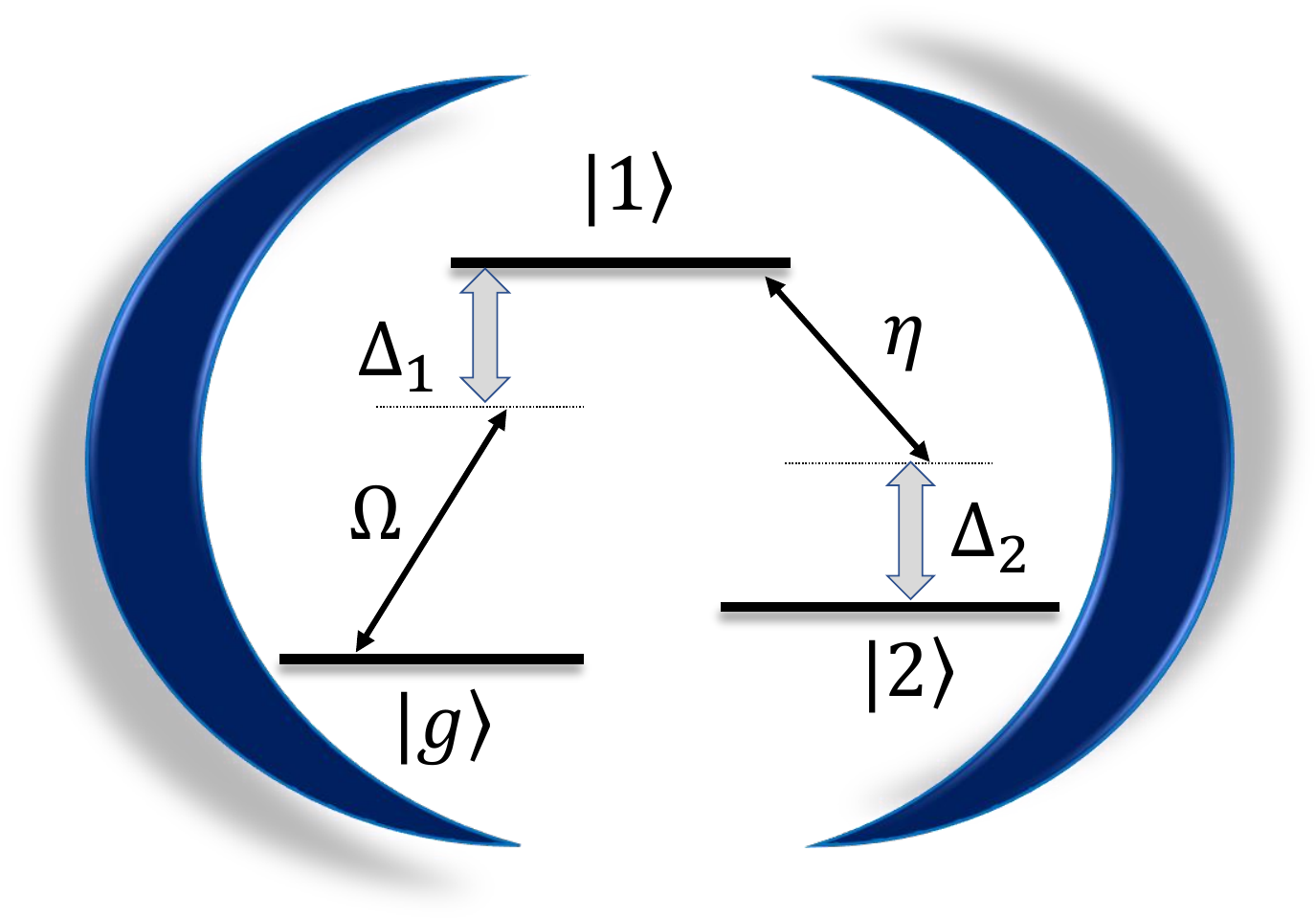} 
	\caption{ A typical lambda system inside a cavity is shown. The transition $\ket{g}\leftrightarrow \ket{1}$ is controlled by the Rabi frequency $\Omega(t)$ and the corresponding detuning $\Delta_{1} \;(\omega_{21}-\omega_{l})$. The transition $\ket{1} \leftrightarrow \ket{2}$ happens under cavity the mode $\omega_{c}$ and governed by the detuning $\Delta_{2} \; (\omega_{12}-\omega_{c})$ and the coupling constant $\eta$ .} 
\label{2photon_Raman}
\end{center}
\end{figure}

As an example, we consider a three-level atomic system in the so-called lambda configuration inside a cavity as shown in the Fig. \ref{2photon_Raman}.
The atomic system interacts with a pump mode of frequency $\omega_l$ and a Stokes mode of frequency  $\omega_{c}$ which is the frequency of cavity mode. The Hamiltonian describing  the system can be written as  
\bea
\mathcal{H} & = &  \omega_{g} \sigma_{gg} + \omega_{1} \sigma_{11} + \omega_{2} \sigma_{22} +  \omega_{c} a^{\dagger} a + \Omega(t)\exp{(-i \omega_{l} t)} \sigma_{1g} \nn \\
& + &  \eta \exp{(-i \omega_{c} t)}\; a\sigma_{12} + \mathrm{H.c.}
\label{Hami4}
\eea
By using the rotating frame and the rotating wave approximation the Hamiltonian of the system reads
\bea
\mathcal{H}_{I}(t) &=& \Omega(t) \exp{(i\Delta_{1} t)} \sigma_{1g} +  \eta  \exp{(i\Delta_{2} t)} a \sigma_{12} + \mathrm{ H.c.} 
\label{Hami5}
\eea
To establish an effective two-level Raman transition  between $\ket{g} \leftrightarrow \ket{2}$, we need to eliminate the state $\ket{1}$ from the effective dynamics. Since in this case $N=2$, therefore, in the Eq. (\ref{main_effec}), we have $\tilde{\Omega}_{0}^{(N-2)} (t)  = \Omega(t); \; \mathcal{C}_{0}^{(1)}  = -\frac{1}{\Delta_{2}} -\frac{1}{\Delta_{1}};\; \tilde{\Delta}_{0}^{(1)}  = -\Delta_{2} + \Delta_{1} $. By inserting these values into the Eq. (\ref{main_effec}), we get the effective Hamiltonian
\bea
\mathcal{H}_{\text{eff}}(t) &=&  -\eta  \Omega(t) \bigg(\frac{1}{\Delta_{1}} +\frac{1}{\Delta_{2}}\bigg) \exp{(i(\Delta_{1}-\Delta_{2})t)} a^{\dagger} \sigma_{2g} + \mathrm{H.c.},
\eea
which can be  obtained following the Jame's effective Hamiltonian theory of two-photon transition [\cite{james2}].
\label{three_photon}
\subsection{The effective Hamiltonian  for a three-photon transition using generalized James' approach}
\label{3photon}
\begin{figure}[hbt!]
	\begin{center}
	\captionsetup[subfigure]{labelformat=empty}
	\includegraphics[width=7.0cm,height=5.cm]{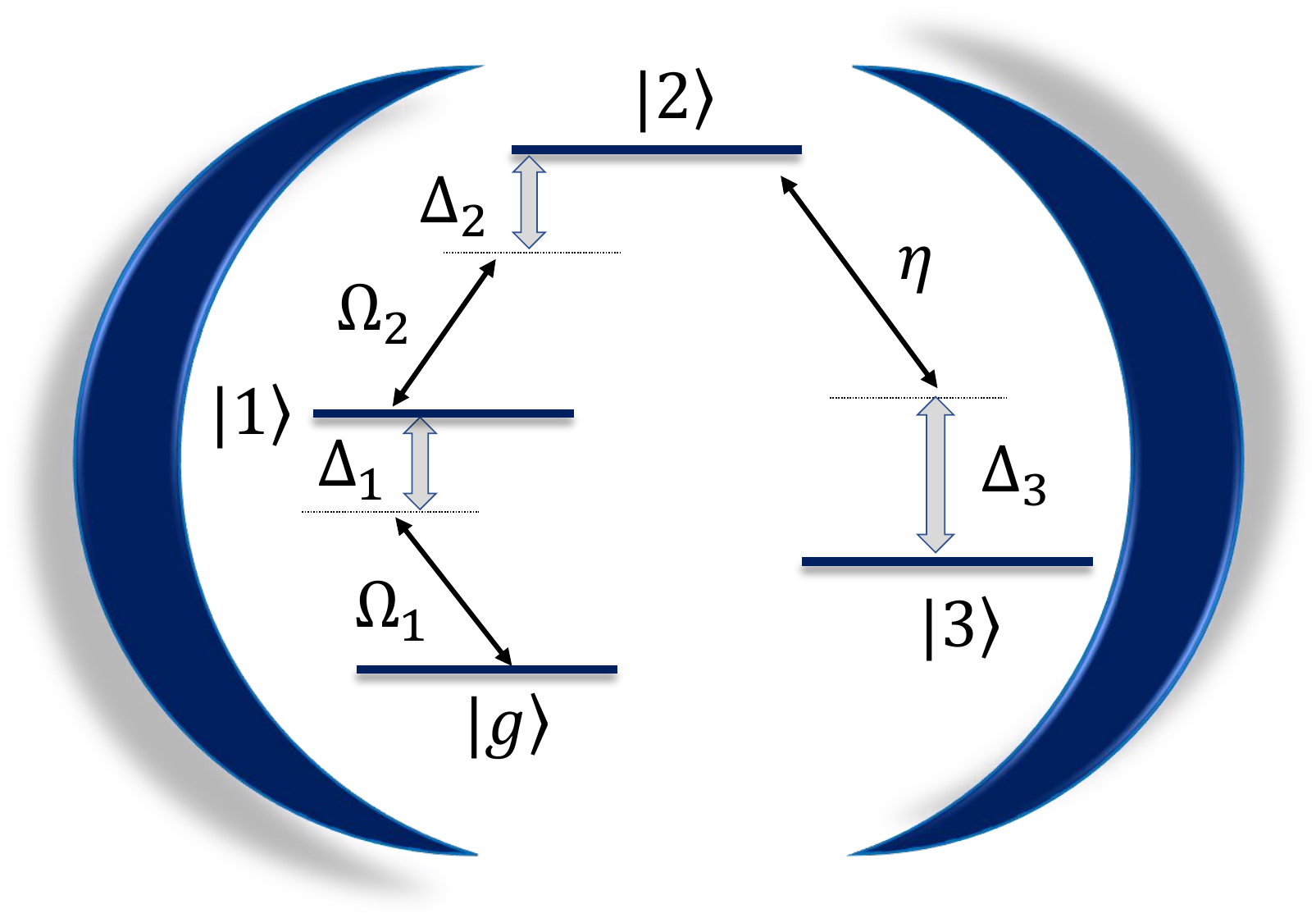} 
	\caption{ A typical three-photon hyper Raman transition inside a single mode cavity of frequency $\omega_{c}$ is displayed. The detuning $\Delta_{1} (\Delta_{2})$ and the Rabi frequency $ \Omega_{1}(t)(\Omega_{2}(t))$ facilitate the transition $\ket{{g}}\leftrightarrow\ket{{1}}(\ket{{1}}\leftrightarrow\ket{2})$. The coupling constant $\eta$ and detuning $\Delta_{3}$ control the transition $\ket{{2}}\leftrightarrow \ket{{3}}$.} 
\label{3photon_Raman}
\end{center}
\end{figure}

To realize the  three-photon transition process,  we consider a four-level atom inside a cavity as shown in the Fig. \ref{3photon_Raman}. We assume the energies of the states $ \ket{g}, \; \ket{1},\;\ket{2}, \;  \ket{3}$ are $E_{g},\;  E_{1},\;  E_{2}$ and $ E_{3}$ respectively. To establish an effective three-photon transition between $\ket{g} \leftrightarrow \ket{3}$, firstly we employ the generalized James' effective Hamiltonian method [\cite{gjames}]. In the interaction picture, the Hamiltonian of the system reads
\bea
\mathcal{H}_{I}(t) &=& \Omega_{1}(t) \exp{(i\Delta_{1} t)} \sigma_{\text{1g}} +  \Omega_{2}(t) \exp{(i\Delta_{2} t)} \sigma_{\text{21}} + \eta \exp{(i\Delta_{3} t)} a \sigma_{23} + \mathrm{ H.c.}
\label{Hami9}
\eea
From the Eqs.(\ref{gen_james_eff}) and (\ref{Hami9}), we can consider  $\omega_1= \Delta_1, \; \omega_2 = \Delta_2, \; \omega_3= \Delta_{3} $ and the corresponding $ \Lambda_{\alpha} \; (\alpha=1,2,3)$ are of the form as $\Lambda_1 = \Omega_{1} \sigma_{1g}$, $\Lambda_2 = \Omega_{2} \sigma_{21}$ and $\Lambda_3 = \eta  a \sigma_{23}$. By utilizing the third-order James’ effective Hamiltonian method (\ref{gen_james_eff}) and the Hermiticity condition used in (\ref{gen_james_con}), we arrive at
\bea
\mathcal{H}_{\text{eff}}^{(3)}(t) = \frac{\eta \Omega_{1}(t) \Omega_{2}(t)}{\Delta_{2} \big(\Delta_1 +\Delta_2\big)} \; a^{\dagger} \sigma_{3g} +\mathrm{H.c.} +\text{Stark shifts.}
\label{Hami10}
\eea
This establishes the effective dynamics between the states $\ket{g}$ and $\ket{3}$ at resonance.  However, we show that our approach based on the Markov approximation approach produces effective dynamics at off-resonance as well as at the resonance. 
\subsection{The effective Hamiltonian using our Markovian approximation approach}
To this end, we revisit to the aforesaid effective dynamics for three-photon hyper Raman  transition process using our Markovian approximation approach.  In this case, $N = 3$, and in the Eq. (\ref{main_effec}), we have \;  $\tilde{\Omega}_{0}^{(1)}  = \Omega_{1}^{*} \Omega_{2}^{*}$, $\mathcal{C}_{0}^{(2)}  = \frac{1}{\Delta_{3}} \mathcal{C}_{0}^{(1)} - \frac{1}{\Delta_{1}}\mathcal{C}_{1}^{(1)} = \frac{1}{\Delta_{3}} \left( \frac{1}{\Delta_{2}}  - \frac{1}{\Delta_{1}}\right) - \frac{1}{\Delta_{1}} \left( \frac{1}{\Delta_{3}} - \frac{1}{\Delta_{2}} \right)$,  and $ \tilde{\Delta}_{0}^{(2)}  = \Delta_{3}+\Delta_{2}+ \Delta_{1}$. Therefore, the effective Hamiltonian (\ref{main_effec}) will take the following form  
\bea
\mathcal{H}_{\text{eff}}(t)  = \eta \Omega_{1}^{*} \Omega_{2}^{*} \left( \frac{1}{\Delta_{3}} \left( \frac{1}{\Delta_{2}}  - \frac{1}{\Delta_{1}}\right)- \frac{1}{\Delta_{1}} \left( \frac{1}{\Delta_{3}} - \frac{1}{\Delta_{2}} \right)\right) \exp{\left(-i \Delta_{0}^{(2)} t \right)} a^{\dagger} \sigma_{g3} + \text{H.c}.
\label{3photon_markov}
\eea
Now, at the resonance we can put $\Delta_{1} + \Delta_{2} + \Delta_{3} = 0 $. As the detunings  $\Delta_{1}$ and $\Delta_{2}$ are large enough, we can neglect the first term within the first bracket in the Eq. (\ref{3photon_markov}). Therefore, we have the effective Hamiltonian at resonance as
\beq
\mathcal{H}_{\text{eff}} (t)  = \frac{\eta \Omega_{1}^{*}(t) \Omega_{2}^{*}(t)} {\Delta_{2} (\Delta_{1}+\Delta_{2})} a^{\dagger} \sigma_{g3} + \text{H.c}.,
\eeq
which is obtained above  (\ref{Hami10}) employing the generalized James' effective  Hamiltonian approach.
\section{Summary and discussion}
\label{con}
We have proposed an approach to derive the effective Hamiltonian under the adiabatic elimination process for multiphoton transition in atom in the presence of classical laser fields and the cavity mode interaction.  By calculating the equations of motion for the transition operators in the  Heisenberg picture and solving it under Markov approximation, we have obtained  the effective dynamics for the system.  We have compared our procedure with the James' and generalized James' effective Hamiltonian approaches with a few examples. From the above analysis it is evident that our approach works well at off-resonance and as well as at resonance.
Unlike the generalized James' approximation approach, our approach can be utilized to get rid of cavity modes from the effective dynamics. For instance, the implementation of  holonomic quantum gates  demands that the  exclusion of cavity modes from the effective Hamiltonian  in order to make the gate operations resilient to the cavity decay [\cite{zhao2018}]. Our approach could be convincingly adopted to  implement the multi-qubit holonomic quantum gates in quantum computation.

\section*{Acknowledgement}
I would like to thank M. Sanjay Kumar for his helpful comments and encouragement.  I acknowledge the financial support from DST (India) through the INSPIRE Fellowship Programme.

\addtocounter{section}{0}
\setcounter{equation}{0}
\numberwithin{equation}{section}
\renewcommand{\thesection}{B}
\addtocounter{section}{0}
\setcounter{equation}{0}
\numberwithin{equation}{section}



\end{document}